\documentclass[final,5p,times]{elsarticle}

\usepackage{epsfig}
\usepackage{latexsym}
\usepackage[colorlinks=true,linktocpage=true,linkcolor=blue,citecolor=blue,allcolors=blue]{hyperref}
\usepackage{url}
\usepackage[utf8]{inputenc}
\usepackage{enumerate}
\usepackage{color}
\usepackage{xcolor}
\usepackage{amsmath,amssymb}

\usepackage[english]{babel}
\hypersetup{
   colorlinks=true, 
  linktoc=all,     
  linkcolor=blue,  
}

\def\be{\begin{equation}}
\def\ee{\end{equation}}
\def\bearr{\begin{eqnarray}}
\def\eearr{\end{eqnarray}}

\journal{Physics Letters B}

\begin{document}

\begin{frontmatter}

 \title{Determination of the $\phi$-meson production process \\ and its absorption cross section via directed flow}

\author[first,second]{Jan Steinheimer}
\affiliation[first]{organization={GSI Helmholtzzentrum f\"ur Schwerionenforschung GmbH},
            addressline={Planckstr. 1}, 
            postcode={D-64291}, 
            city={Darmstadt},
            country={Germany}
}

\affiliation[second]{organization={Frankfurt Institute for Advanced Studies (FIAS), Ruth-Moufang-Str. 1, D-60438 Frankfurt am Main, Germany}}

\author[third,second,fourth]{Tom Reichert}
\affiliation[third]{organization={Institut für Theoretische Physik, Goethe-Universit\"{a}t Frankfurt, Max-von-Laue-Str. 1, D-60438 Frankfurt am Main, Germany}}
\affiliation[fourth]{organization={Helmholtz Research Academy Hesse for FAIR (HFHF), GSI Helmholtzzentrum f\"ur Schwerionenforschung GmbH, Campus Frankfurt, Max-von-Laue-Str. 12, 60438 Frankfurt am Main, Germany}}

\author[third,first,fourth]{Marcus Bleicher}


\begin{abstract}
It is shown that the directed flow of $\phi$-mesons in Au+Au collisions at $\sqrt{s_{NN}}=3$ GeV, is sensitive to the production and the absorption cross section of the $\phi$ in a nuclear medium. This provides a new observable to constrain the in-medium properties of the $\phi$ which is independent of its absolute production rate. STAR data disfavor any significant $\phi$-N absorption in dense nuclear matter and are consistent with a very small cross section of the $\phi$ comparable to the vacuum cross section. The similarity of the $\phi$-meson and proton directed flow also indicates that the $\phi$ is produced in conjunction with a baryon.
\end{abstract}
\end{frontmatter}


\section{Introduction}
The properties of hadronic resonances in relativistic nuclear collisions have been an important field of study as they can carry information on the properties of the QCD medium produced in such collisions \cite{Bass:2000ib,Bleicher:2002dm,Bleicher:2002rx,Bleicher:2003ij,Vogel:2005qr,Vogel:2005pd,Vogel:2007qg,Vogel:2007yu,Vogel:2009kg,Vogel:2010pb,Hirano:2005xf,Steinheimer:2012era,Knospe:2015nva,Steinheimer:2015msa,Steinheimer:2017vju,Knospe:2021jgt}.
During the hadronic phase of such collisions the various produced hadrons and their resonances interact, decay and re-scatter copiously which allows to study scattering processes which cannot be produced in any other experimental setup.
The investigation of resonances also has a long established experimental history with major contributions from experiments at GSI, BNL and CERN \cite{STAR:2004bgh,Markert:2005jv,STAR:2008twt,HADES:2013sfy,Knospe:2015rja,ALICE:2018ewo,ALICE:2022zuc,ALICE:2023ifn,Kozlowski:2024cjw,Rozplochowski:2024pld,NA49:2008goy,NA61SHINE:2019gqe,FOPI:2002csf,HADES:2017jgz,ALICE:2014jbq,E917:2003gec}.
While for example the vector decay of the $\rho$-meson, into a di-lepton, tells us something about the dense phase (see e.g. \cite{Shuryak:1978ij,Koch:1992sk,McLerran:1984ay,Bratkovskaya:1996qe,Rapp:1999ej,HADES:2019auv}), the hadronic decay of the $K^*$-meson allows us to infer the lifetime of the hadronic phase \cite{Chabane:2024crn,Neidig:2025xgr}.
It has also been suggested, that the rescattering of the $f_0$ resonance may even be sensitive to its quark structure \cite{Reichert:2024stb}.

The $\phi$-meson (and its charmed counterpart the $J/\psi$) is of particular interest. In the vacuum, the lifetime of the $\phi$ ($\tau_{\phi}\approx 40 \ \mathrm{fm}/c$) is longer than the expected lifetime of a fireball created in nuclear reactions and the inelastic ${\phi+\mathrm{nucleon} \rightarrow X}$ cross section is considered small \cite{ParticleDataGroup:2024cfk}. This may change in the dense nuclear medium due to interactions. Several studies have tried to measure the absorption cross section of the $\phi$ in nuclear matter by comparing the total production cross section in collisions of protons with nuclei of different mass number $A$ \cite{Hartmann:2012ia,Ishikawa:2004id,CLAS:2010pxs}. 
The measurement presented in these works is based on the assumption, that the $\phi$ production cross section in p+$A$ scattering, in the absence of any further interactions, is simply proportional to some power of the mass number $A$ of the target nucleus. It is found that the measured $\phi$ production in large target nuclei, when scaled to the cross section in p+C, is significantly lower than what is expected from this simple scaling argument. The difference in the production cross section is then attributed to absorption of the $\phi$ in the target nucleus. It was concluded, that the absorption cross section of the $\phi$ in nuclear matter may be rather large, on the order of ${\sigma_{\mathrm{abs}}\approx 20}$ mb. Similar studies were done for the $J/\psi$ at CERN-SPS energies \cite{Kharzeev:1996yx} suggesting also a rather large hadronic $J/\psi$ absorption cross section on the order of 7-8 mb. Further studies of the $J/\psi$ are also envisioned utilizing the enormous collision rate of the FAIR facility. 

With respect to this line of argument we want to point out, that the reduction of the $\phi$ production in large target nuclei may simply be because the original production of the $\phi$ may show a different scaling with mass number $A$ and not because it is first produced and then absorbed. It was indeed shown in one of our previous publications \cite{Steinheimer:2015sha} that the observed reduction of the $\phi$ production on different target nuclei as observed by the ANKE experiment can very well be reproduced without any additional in-medium effect on the $\phi$ absorption cross section. In the present paper we want to present an alternative measurement which is able to distinguish between the two effects of reduced production vs. absorption inside dense matter.

\begin{figure*} [t]
    \centering
    \includegraphics[width=\textwidth]{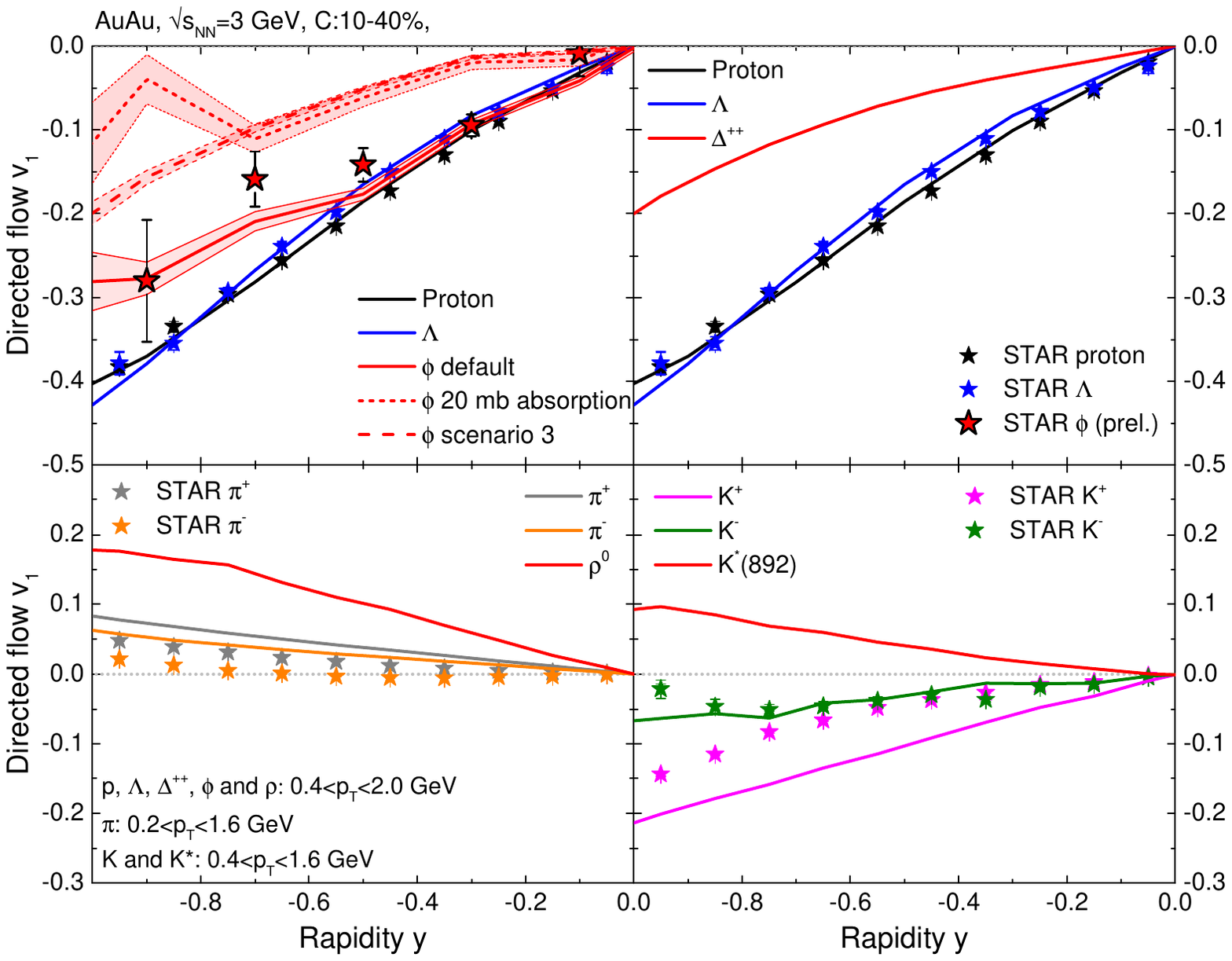}
    \caption{[Color online] Directed flow of various hadrons as function of rapidity in the center of mass frame of the Au+Au collision at $\sqrt{s_{NN}}=3$ GeV for 10-40$\%$ most central reactions. Stable hadron flow is compared to STAR data. Different resonances are shown as red lines in every plot. For the $\phi$-meson, we compare results with the default absorption (red solid line) with a constant 20 mb absorption cross section (red dashed line). STAR data are taken from \cite{STAR:2021yiu,poster}.}
    \label{fig:flow}
\end{figure*}

\begin{figure} [t]
    \centering
    \includegraphics[width=0.5\textwidth]{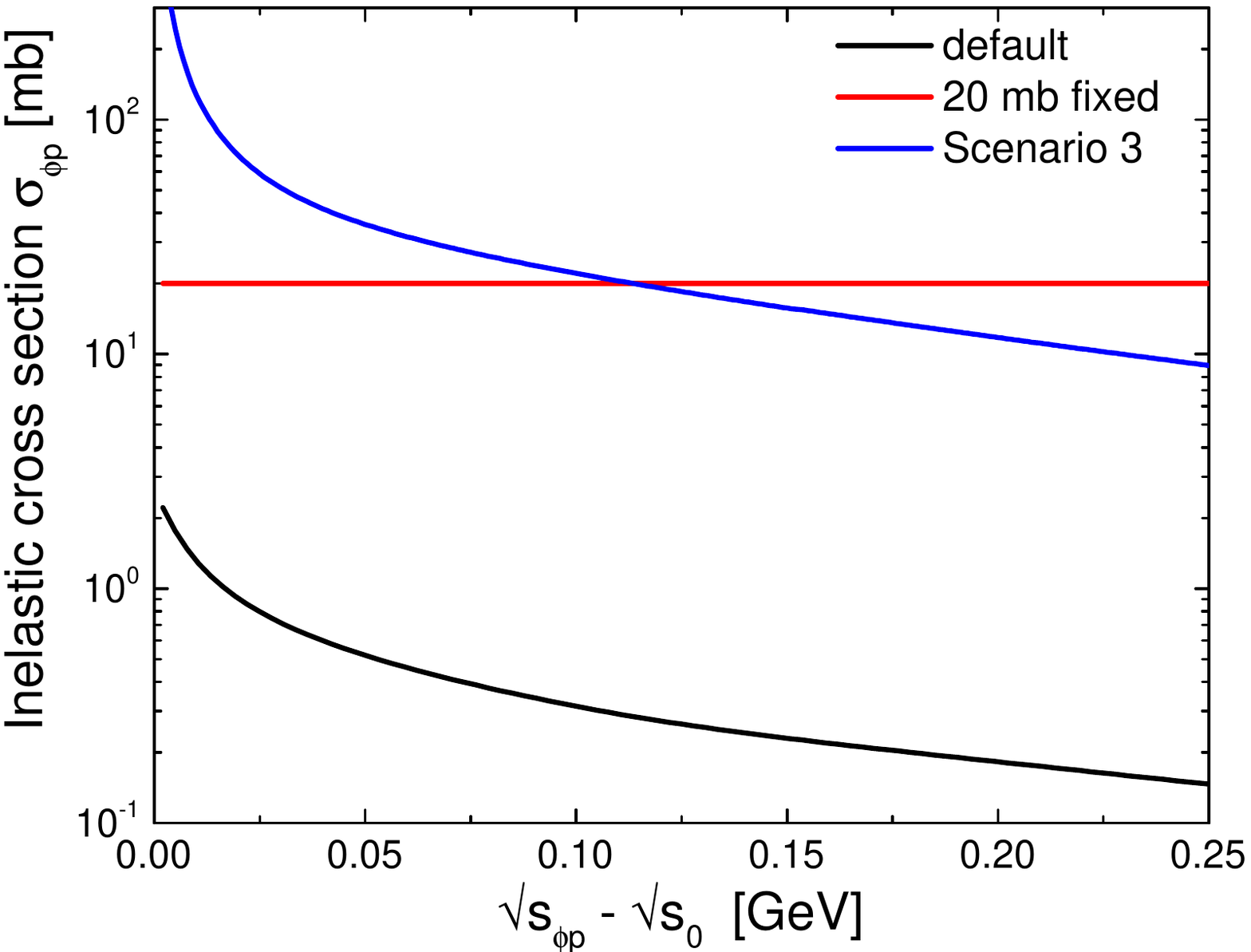}
    \caption{[Color online] Total inelastic cross section of  ${\phi+p\rightarrow N^*}$ as function of the invariant mass minus the sum of masses of the two hadrons. The black line corresponds to the cross section consistently calculated from the inverse of the production reaction, i.e. via detailed balance as implemented in the model.}
    \label{fig:phi}
\end{figure}
The production of the $\phi$, including full in-medium properties and cross sections, has also been discussed within the PHSD model recently \cite{Song:2022jcj} to better understand its production mechanism close to its elementary threshold energy.

In this work we will show how the in-medium absorption of the $\phi$-meson can be inferred from flow data of the STAR experiment in a measurement that is independent of the total production cross section.

The idea is that, after the $\phi$ has been produced in the nuclear reaction, if it has only a very small in-medium cross section it will leave the system almost undisturbed and therefore inherit the flow of the particle it was produced with. This would be the proton in case of the STAR measurement. If the $\phi$ had a significant absorption and/or regeneration cross section in the medium, the directed flow would be significantly modified as we will see for other short lived resonances.

\section{UrQMD and resonances}

We will employ the UrQMD transport model (v4.0) \cite{Bass:1998ca,Bleicher:1999xi} to simulate the dynamics of Au+Au reactions at ${\sqrt{s_{NN}}= 3}$ GeV and compare the results to recent (preliminary) STAR data. The STAR experiment has published the directed flow 
\begin{equation}
{v_1(y)=\left<p_x / p_T \right>(y)}
\end{equation}
of different identified hadrons, where $y$ is the rapidity in the center of mass frame of the collision, $p_x$ is the momentum in x-direction and $p_T$ is the  transverse momentum of a hadron. The averaging is done in a given rapidity bin $y$ and $p_T$ acceptance of the experiment.\\

The production of hadrons in the UrQMD transport model
\cite{Bass:1998ca,Bleicher:1999xi,Bleicher:2022kcu} proceeds through different channels: The excitation and de-excitation (decay) of hadronic resonances, of a string and the annihilation of a particle with its anti-particle. The probabilities of the different processes are governed by their reaction cross sections. These cross sections serve as input for the model and are taken, whenever possible, from experimental measurements of
elementary (binary) collisions. 
The basic input for our model calculations are the elementary hadron scattering cross sections in the vacuum. For some reactions, these are known and explicitly measured and for others they are inferred. For example, the $K^+ + p$ cross section has been measured explicitly and this is what is used as input in the UrQMD simulations. The $\phi+p$ cross section is not explicitly measured. What is used in the UrQMD model is calculated from detailed balance using the $\phi$ production cross section in p+p collisions \cite{Steinheimer:2015sha} which is what we refer to as the default cross section.
The $\phi$-meson production at near- and subthreshold energies is treated, in analogy to $\rho$-meson production via the decay of a heavy resonance $N^* \rightarrow N + \phi$ \cite{Steinheimer:2015sha}. The $\phi$ absorption cross section is in turn coupled via detailed balance to the partial decay width. It is important to note that this process always couples the $\phi$ to its parent baryonic resonance and one would assume that the meson inherits the flow properties from the baryons, i.e. following the baryon flow. This is different from what one would expect if the $\phi$ is produced e.g. from recombination of a pair of $s + \overline{s}$ quarks.

In order to describe the flow of bulk matter at STAR fixed target energies, the inclusion of a realistic equation of state (EoS), via the QMD potentials, is necessary. In the following we will employ the most recent version of the momentum dependent potentials from the CMF model which have been shown to describe proton and hyperon flow as well as strange hadron production yields as measured by the HADES and STAR experiments \cite{Steinheimer:2025jll,Steinheimer:2024eha,Reichert:2025rnw}.

The flow of stable hadrons is extracted at their time of last interaction, i.e. their kinetic freeze-out point. For observables involving decaying resonances, an observable resonance is defined by following their decay products throughout the entire systems evolution, and if none of the decay products undergoes any rescattering the resonance is considered reconstructable and can be used in our flow analysis. This method is well established and is compatible with the experimentally employed invariant mass analysis \cite{Bleicher:2002dm,Reichert:2019lny,Chabane:2024crn}.

\section{Results}

Figure \ref{fig:flow} shows the directed flow as function of rapidity, for a variety of stable hadrons and resonances, in 10-40$\%$ most central Au+Au collisions at $\sqrt{s_{NN}}=3$ GeV.

The proton (black solid lines) and hyperon (blue solid lines) flow is compared to experimental data from the STAR collaboration (black and blue star symbols) \cite{STAR:2021ozh,STAR:2021yiu}. The model describes these data very well, indicating that the bulk evolution and bulk flow is well described by UrQMD. 

The mesons are shown in the lower panels of figure \ref{fig:flow} as orange and gray solid lines (positively and negatively charged pions) and as green and magenta solid lines ($K^-$ and $K^+$), whereas the STAR data \cite{STAR:2021yiu} is shown as colored symbols. Both pions and (anti-)Kaons are also reasonably well described, although the pion flow in the model is slightly too large and Kaon flow slightly too small.
 The Kaon, unlike the $\phi$, cannot be absorbed through scattering with a nucleon due to the conservation of strangeness and only changes its momentum. This means, for the directed flow of Kaons, subtleties in the angular distribution of Kaon-nucleon scattering may also play a role or may be related to the fact, that UrQMD-v4.0 does not include explicit meson potential interactions which can become relevant for the Kaons. This means that more effort has to be put into understanding the Kaon scattering (also in the medium) which is however not the topic of the present paper.

The red lines in all four panels show the resulting flow of reconstructable resonances, the $\phi$ (top left), $\rho^0$ (bottom left), $\Delta^{++}$ (top right) as well as the $K^*$ (bottom right). As one can clearly see, all resonances, except the $\phi$-meson, deviate significantly from the stable hadron flow and show a reduction of emission in the direction of the spectators. This is due to the complex dynamics of decay and regeneration that these resonances undergo due to their short lifetime and often large cross section of the resonance and its daughter particles with the nucleons. 

The $\phi$-meson on the other hand, is either totally absorbed or not, making it a much cleaner probe. It either leaves the system completely undisturbed once it has been produced on the nucleon resonance or is absorbed by an elastic rescattering with a nucleon.

\subsection{$\phi$ absorption}
To quantify what would happen if the $\phi$ had indeed a significantly larger absorption cross section (as was suggested e.g. in \cite{Hartmann:2012ia}), we compare three different scenarios:
\begin{enumerate}
\item Scenario 1 (Default): This is the default scenario for the $\phi$ absorption where the ${\phi+N\leftrightarrow N^*}$ cross section is constraint from detailed balance and the inverse production cross section in elementary p+p reactions as described in \cite{Steinheimer:2015sha}. The resulting total inelastic cross section is shown as black line in figure \ref{fig:phi}. Even for very small relative momenta the absorption cross section does not exceed 2 mb.  
\item Scenario 2 (20 mb fixed): In this scenario we simply fix the ${\phi+N\leftrightarrow N^*}$ cross section to a constant value of 20 mb, in line with conclusions drawn from \cite{Hartmann:2012ia}. This is shown as red line in \ref{fig:phi}. This fixed cross section violates detailed balance as the back reaction is not increased accordingly. One may argue that this could be the proper treatment to compare to the cross section extracted in \cite{Hartmann:2012ia} as the total absorption would also include any additional production in the medium.
\item Scenario 3: The $\phi$ production cross section has been increased by a factor of 100 consistently with the absorption cross section to ensure detailed balance. This leads to a energy dependent cross section which is also shown as blue line in \ref{fig:phi}.
\end{enumerate}

 The resulting directed flow for these three scenarios is shown as red lines (solid and dashed) in the top left panel of figure \ref{fig:flow}. The increased absorption has a significant impact on the observed directed flow of the $\phi$ and, similar to the other resonances, leads to a reduction of the flow for the $\phi$ in the direction of the spectators. The strong deviation is also observed to be independent of the actual production cross section. 

When comparing to the (preliminary) data from the STAR experiment \cite{poster}, the calculation with a standard small absorption cross section, not modified in the medium, best describes the data.

Despite the large statistical errors in the experimental data for large rapidities, the deviations from the data with small errors can be seen clearly. As one can also see, the increase of the directed flow towards large rapidities, is not a result of limited statistics but rather a result of the large absorption in the spectator. The effect is largest if a constant absorption cross section of 20mb is used.

\section{Conclusion}

We have shown that the directed flow of the $\phi$-meson, as measured by the STAR collaboration in peripheral Au+Au collisions at $\sqrt{s_{NN}}= 3$ GeV, is a sensitive probe of the absorption cross section of the $\phi$-meson in dense nuclear matter. In addition, this observable has the advantage that it is insensitive to the total production probability, which makes it more reliable than measurements of $\phi$ absorption in p+A collisions.
The similar directed flow of the $\phi$ and the protons indicate that the $\phi$ inherits its flow from the baryon it was produced on, which supports the idea of $\phi$ production in a secondary scattering of heavy baryonic states rather than early production, i.e. in a color string.

We propose to extend this study to the directed flow of the hidden charm $J/\psi$, and therefore its absorption in nuclear matter, which allows to validate the surprisingly large $J/\psi$ absorption cross section extracted in previous studies. The CBM@FAIR experiment is in a prime position to provide these unique data, due to its high luminosity.

\section*{Acknowledgments}
The authors thank Nu Xu for fruitful discussions about the $\phi$-meson flow.
The computational resources for this project were provided by the Center for Scientific Computing of the GU Frankfurt and the Goethe--HLR and GSI green cube.


\end{document}